\documentclass[fullpage,a4paper]{article}

\usepackage{amsmath}
\usepackage{amssymb}
\usepackage{amsthm}
\usepackage{bm}
\usepackage{authblk}
\usepackage{graphicx}
\usepackage{csquotes}
\usepackage{todonotes}
\usepackage{listings}
\usepackage{url}
\usepackage{hyperref}

\newcommand{\fig}[1]{Figure~\ref{fig:#1}}
\newcommand{\hyp}[1]{Assumption~\ref{hyp:#1}}

\newtheorem{hypothesis}{Assumption}

\newcommand{\beginsupplement}{%
  \setcounter{table}{0}
  \renewcommand{\thetable}{S\arabic{table}}%
  \setcounter{figure}{0}
  \renewcommand{\thefigure}{S\arabic{figure}}%
}

\title{Visual explanation of country specific differences in
  Covid-19 dynamics.}
\author{Nils Bertschinger \thanks{bertschinger@fias.uni-frankfurt.de}}

\begin{document}

\maketitle

\begin{abstract}
  This report provides a visual examination of Covid-19 case and death
  data. In particular, it shows that country specific differences can
  too a large extend be explained by two easily interpreted
  parameters. Namely, the delay between reported cases and deaths and
  the fraction of cases observed. Furthermore, this allows to lower
  bound the actual total number of people already infected.
\end{abstract}

\section{Introduction}

The unfolding COVID-19 pandemic requires timely and finessed
actions. Policy makers around the globe are hard pressed to balance
mitigation measures such as social distancing and economic
interests. While initial studies \cite{imperial1} predicted millions
of potential deaths never findings hint at a much more modest outcome
\cite{Lourenco2020.03.24.20042291,imperial2}. Especially the case
fatality rate (CFR) and the number of unobserved infections are
crucial to judge the state of the pandemic as well as the
effectiveness of its mitigation. Yet, there estimates are plagued with
high uncertainties as exemplified in the quick revisions even from the
same institution \cite{imperial1,imperial2}

Most studies are based on elaborate epidemic modeling either using
stochastic or deterministic transmission dynamics. Especially, the
susceptible-infected-recovered (SIR) model \cite{Newman} forms a
basic building block and has been extended in several directions in
order to understand the dynamics of the ongoing Covid-19 pandemic
\cite{arxiv:2002.07572,arxiv:2004.01105,10.1126/science.abb3221,https://www.medrxiv.org/content/10.1101/2020.02.27.20028639v2}.
In this context, it has not only been compared with more
phenomenological growth models
\cite{https://doi.org/10.1101/2020.03.12.20034595}, e.g. logistic
growth, but also been used to quantify the effectiveness of quarantine
and social distancing \cite{arxiv:2002.07572,arxiv:2004.01105}.
E.g. social distancing, can be easily included by replacing the
infection rate parameter with a function allowing it to change over
time. \cite{arxiv:2004.01105} assumes one or several (soft) step
functions where the infection rate drops in response to different
measures after these had been implemented.

Such detailed modeling is required in order to capture and forecast
temporal dynamics of the epidemic spreading. Yet, substantial care is
needed as to which parameters can be learned from the data and which
cannot. Indeed, I show here that SIR type models -- and others
exhibiting similarly flexible growth dynamics -- are non-identified with
respect to the CFR and the fraction of observed infections.
Instead, a direct visual exploration of the data leads to valuable
insights in this regard. In particular, much of the variability
relating reported case and death counts can be explained by two easily
interpreted parameters. Furthermore, based on three simple assumptions
a lower bound on the number of actual infections, including observed
and unobserved cases, can be obtained. In turn, confirming recent
estimates without the need of complex and maybe questionable modeling
choices.

\section{Data exploration}

Covid-19 data are published by several sources, most notably the John
Hopkins university and the European Center for Decease Prevention and
Control (ECDC). Here, data from ECDC as available from
\url{https://opendata.ecdc.europa.eu/covid19/casedistribution/csv} are
used.

\begin{figure}
  \includegraphics[width=0.495\textwidth]{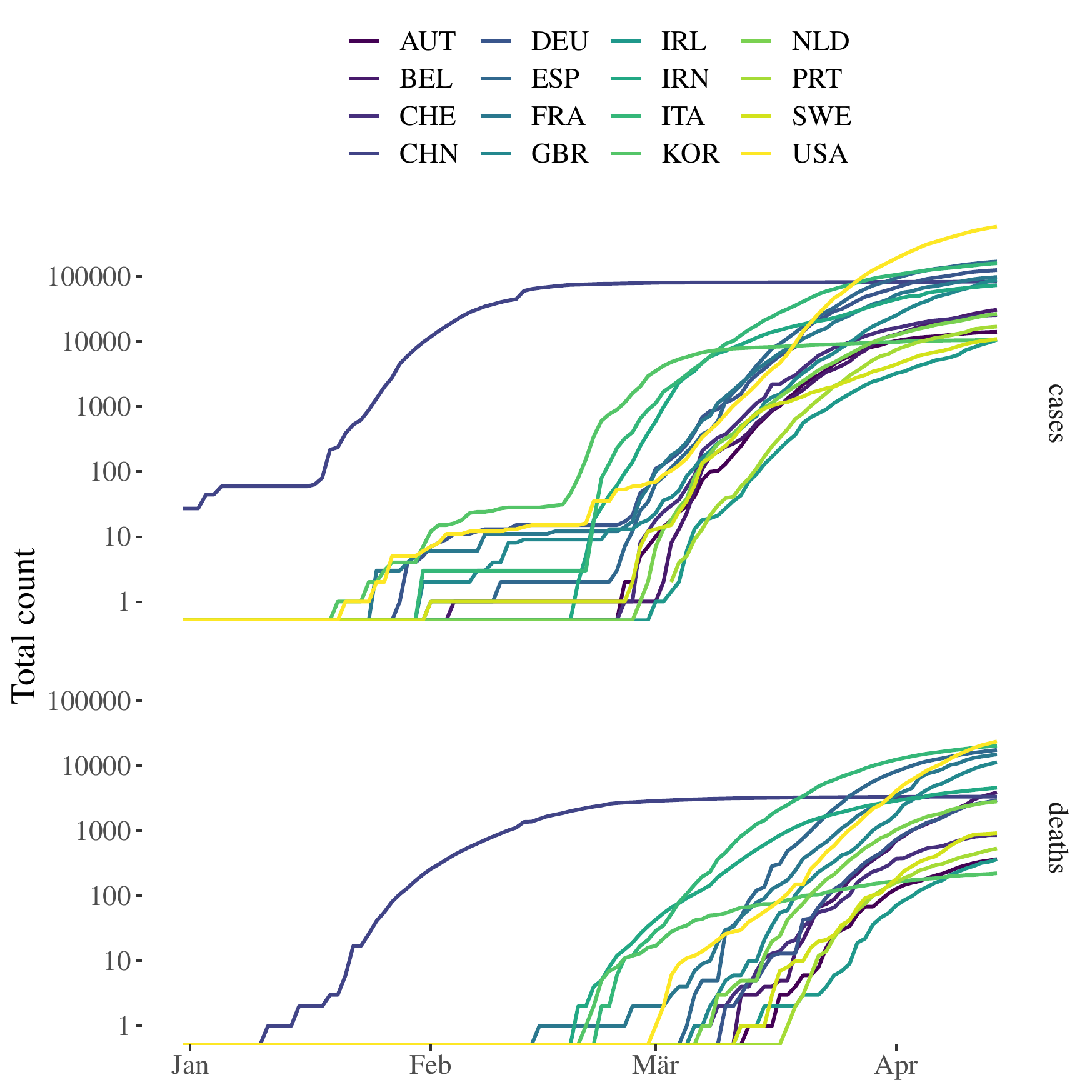}
  \includegraphics[width=0.495\textwidth]{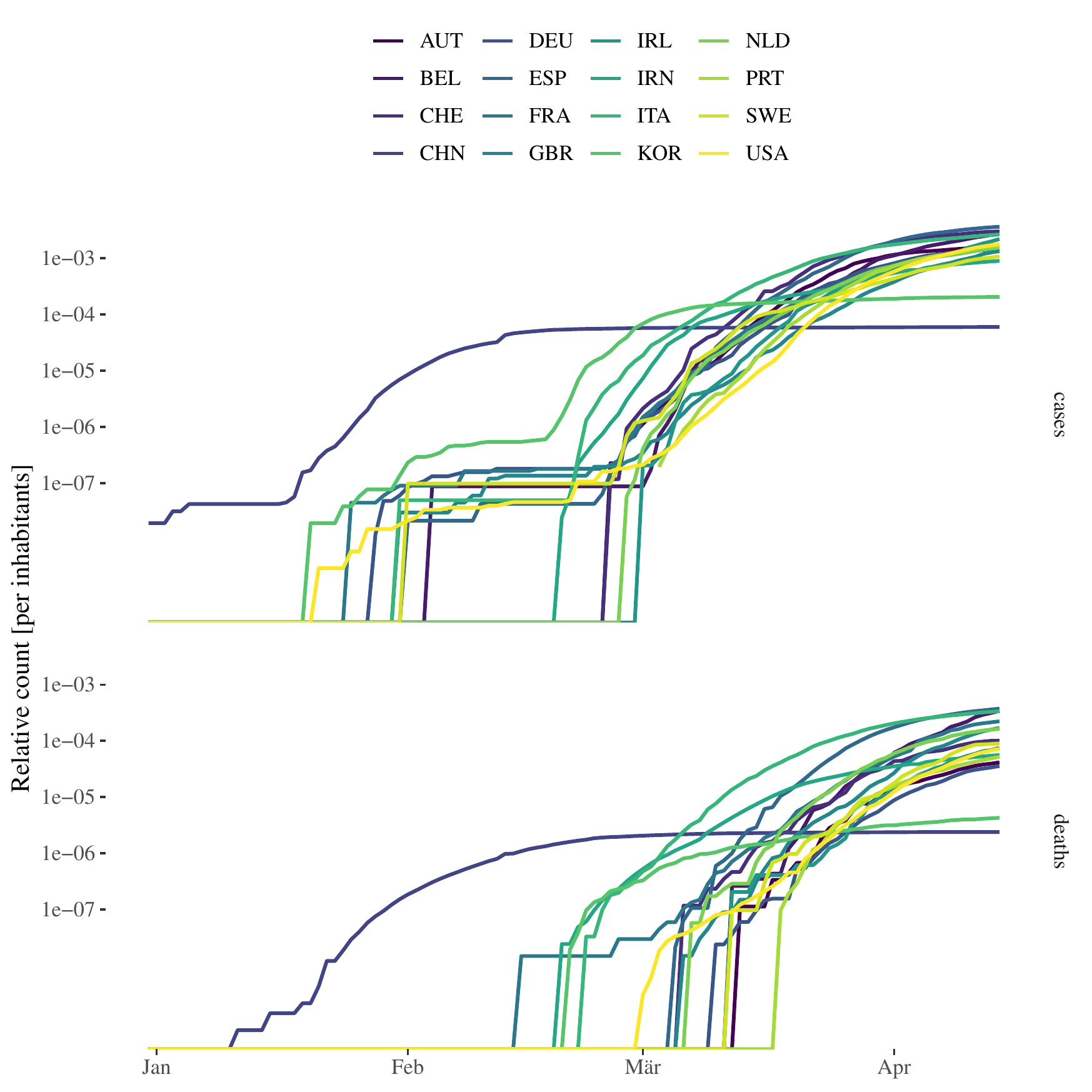}
  \caption{\label{fig:raw_data} Case and death counts of selected
    countries. Both in absolute (left) and relative (right), i.e. per
    inhabitants, terms.}
\end{figure}
\fig{raw_data} shows the total cumulative case and death counts of
selected countries. These countries are among the eight most effected
countries in terms of absolute and relative deaths\footnote{In
  addition, South Korea is included as its numbers are commonly
  considered of high quality.}. In the following, I will focus on
relative counts as these are arguably more meaningful when comparing
different countries -- which could differ widely in terms of
population size.

\begin{hypothesis}
  \label{hyp:count}
  Death counts are more reliable than case counts.
\end{hypothesis}

\begin{figure}
  \includegraphics[width=1\textwidth]{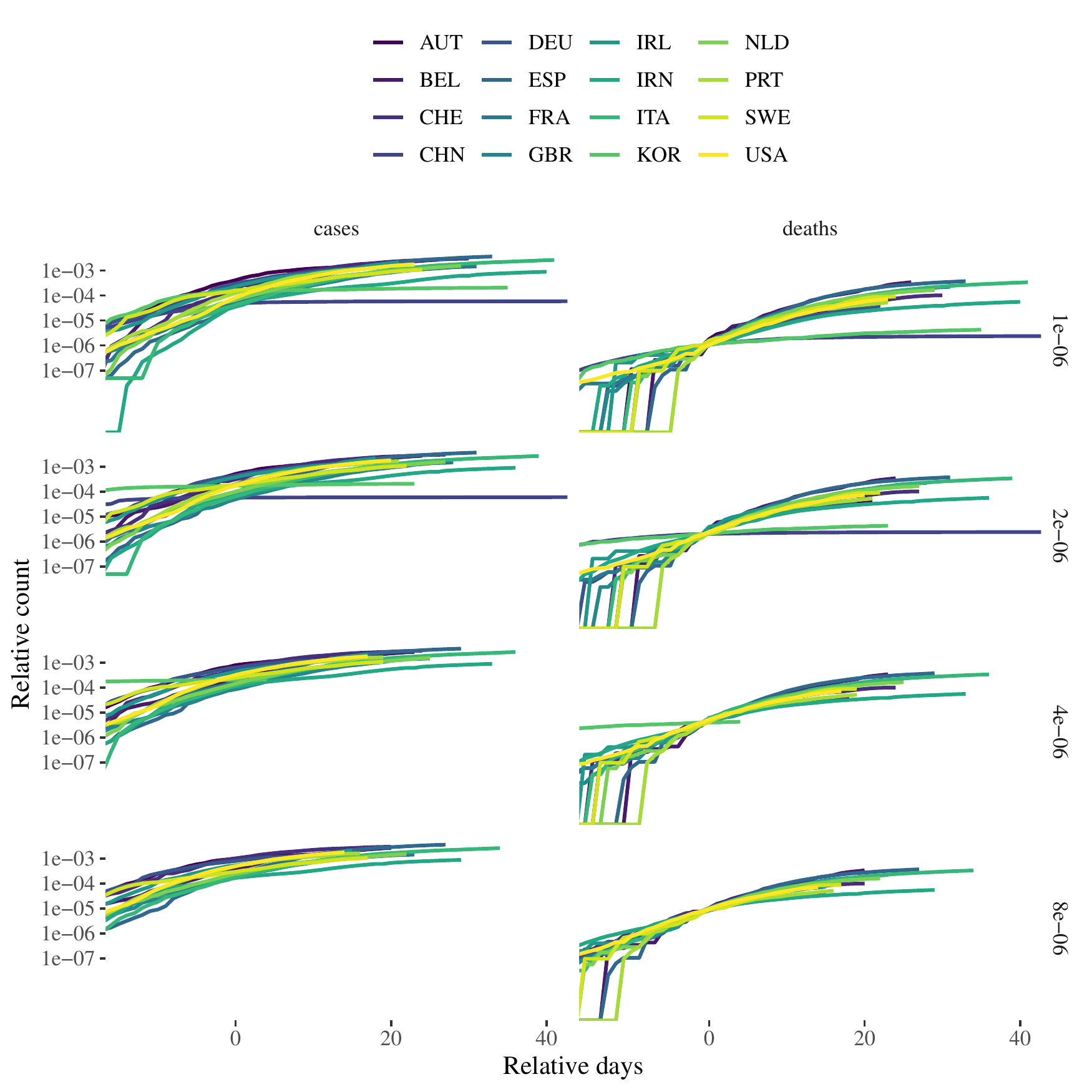}
  \caption{\label{fig:aligned_data} Relative case and death counts of
    selected countries. Dates are aligned relative to the first day
    that relative death counts exceed one (top) or ten (bottom) per
    million respectively.}
\end{figure}

By \hyp{count} analysis will start from relative cumulative death
counts $d_t$ in the following\footnote{Similarly, relative cumulative
  case counts are denoted as $c_t$}. Furthermore, in order to
facilitate country comparisons, dates are shifted relative to the
first day that relative death counts exceed a threshold $\theta$ of
$1, 2, 4$ or $8$ deaths per million inhabitants respectively, i.e. $t
= 0$ is defined such that $d_t \geq \theta$ for $t \geq 0$ and $d_t <
\theta$ for $t < 0$. \fig{aligned_data} shows the resulting time
course of relative case and death counts. Aligning dates in this
fashion shows that several countries exhibit similar time courses,
e.g. Belgium and Spain or China and South Korea. As shown in the
supplementary \fig{scaling} the remaining country specific differences
can be explained by differences in growth rates.  Re-scaling time
according to the estimated doubling time indeed leads to a data
collapse as complete as often observed in physical systems exhibiting
scaling laws \cite{stanley99}.

Here, these differences in the precise temporal dynamics of epidemic
growth are not required.  Instead, the relation between relative death
and case counts is considered. While relative death counts exhibit
similar time courses the corresponding relative case counts $c_t$ are
more variable when aligned in the same fashion, i.e. relative to the
first day that $d_t$ exceeds a given threshold. As I will argue now,
most of this variability can be explained with two readily
interpretable parameters.

\begin{hypothesis}
  \label{hyp:delay}
  There is a well defined country specific delay between reported
  cases and deaths.
\end{hypothesis}

\fig{aligned_data} suggests that relative case counts are not aligned
as some countries, e.g. Germany, systematically lead the counts
reported in other countries, e.g. Italy. Such a difference could mean
that individuals survive longer, e.g. due to differences in medical
care, until they eventually. It could also just reflect reporting
delays due to bureaucratic reasons. In any case, it is clearly the
case that individuals die not immediately, but some days after they
had been tested positive previously.

\subsection{Case fatality rate}

This delay also needs to be taken into account when estimating the
case fatality rate (CFR). Commonly the CFR is defined as $\mathrm{cfr}
= \frac{d_t}{c_t}$. Not surprisingly this estimate is highly variable
and changes systematically over time, especially at the beginning of
an epidemic. The observation captured in \hyp{delay} also explains the
surprisingly low CFRs initially announced in Austria and Germany where
reported death counts are simply some days older compared to other
countries!

Thus, taking into account that individuals that had been tested
positive will usually not die on the same day but after some delay
$\tau$ (if at all), I define
\begin{align}
  \label{eq:cfr}
  \mathrm{cfr}_{\tau} &= \frac{d_t}{c_{t - \tau}} \; ,
\end{align}
i.e. comparing current death with previous case counts.

\begin{figure}
  \includegraphics[width=0.495\textwidth]{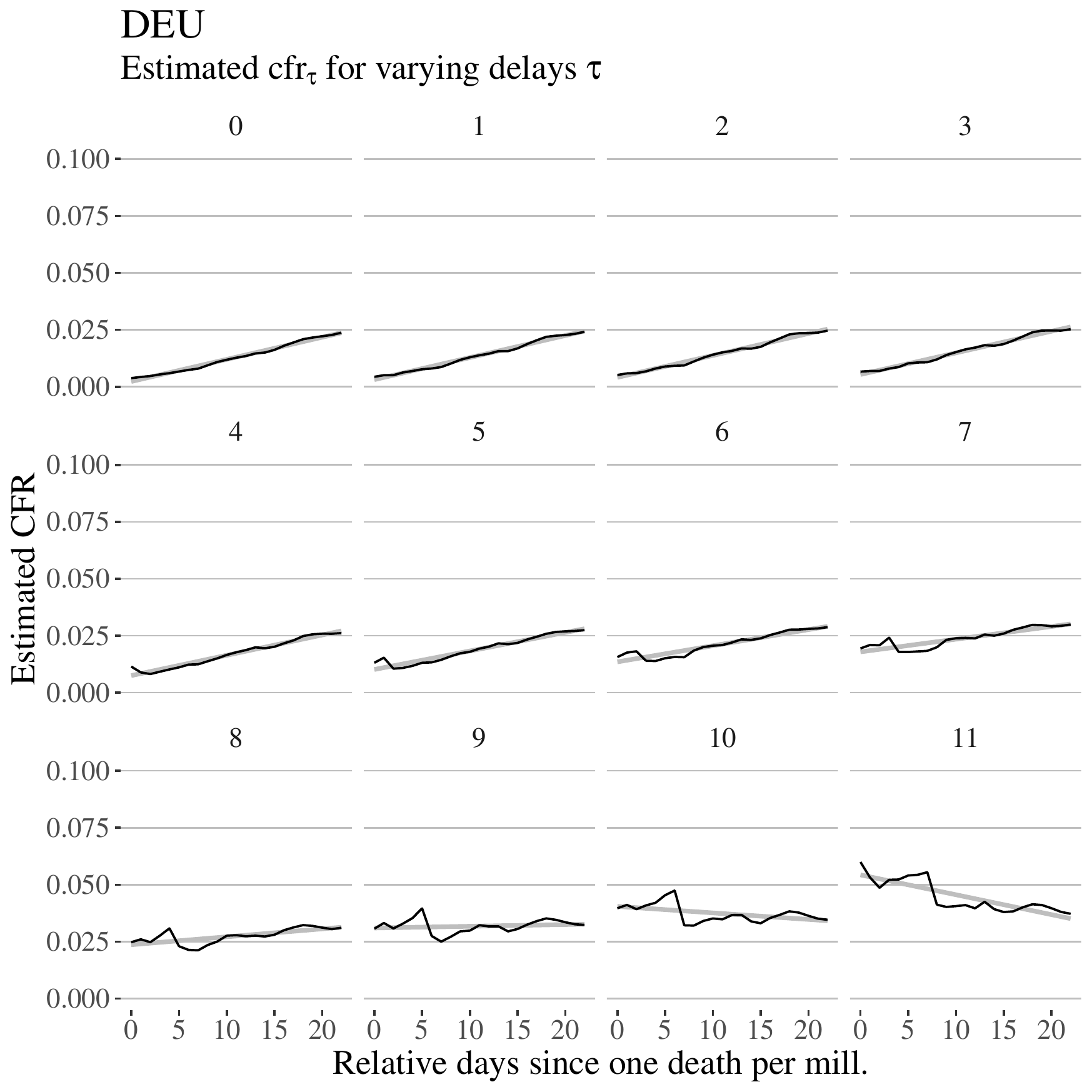}
  \includegraphics[width=0.495\textwidth]{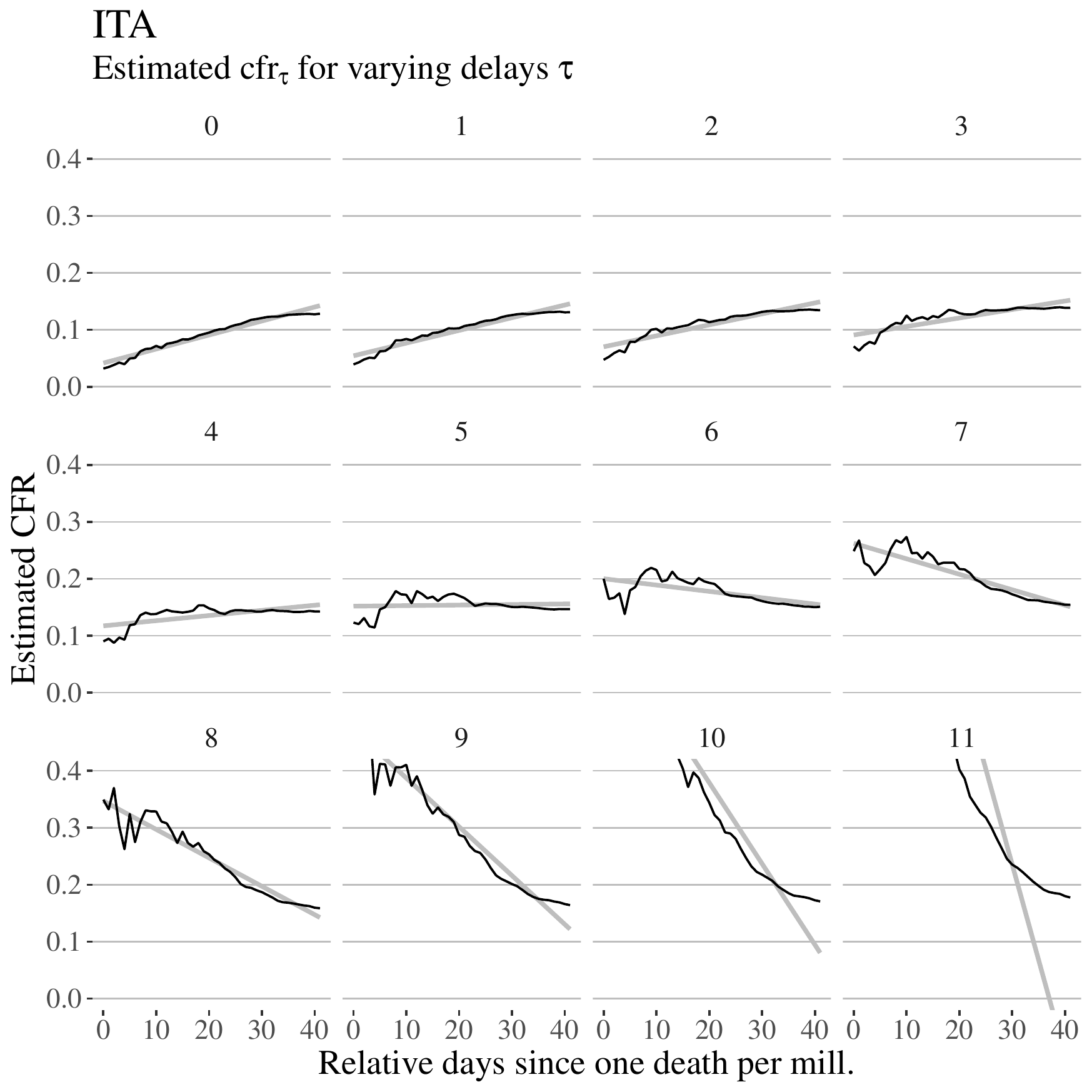}
  \caption{\label{fig:cfr} Estimated CFR $\mathrm{cfr}_{\tau}$ for
    Germany (left) and Italy (right) using different delays of $\tau =
    0, \ldots, 11$ days. Note that in each case, there exists a
    characteristic delay such that estimates are almost constant over
    time. Further note that estimates for all delays will eventually
    converge to the same final value when enough data are available.}
\end{figure}
\fig{cfr} shows the CFRs estimated for Germany and Italy in this
fashion, i.e. for different delays $\tau$. The estimate using $\tau =
0$ rises over time simply reflecting that due to the reporting delay
death counts have not yet caught up with the exponentially growing
case counts. Interestingly, for each country there exists a
characteristic delay at which the estimated CFRs are essentially
constant. Thus, reflecting the hypothesized delay between reported
cases and deaths.

This delay can either be estimated by visual inspection or by fitting
a linear model on each delay and picking the one with minimal absolute
slope\footnote{Just as an ad-hoc algorithm mimicking the visual
  procedure.}. \fig{cfr} shows the delays $\tau$ and corresponding
CFRs $\mathrm{cfr}_{\tau}$, i.e. the median CFR value at this delay,
estimated for each country in this fashion. In order to fully relate
the observed case with death counts an additional, and stronger,
assumption is needed.

\begin{hypothesis}
  \label{hyp:cfr}
  The true case fatality rate is the same for all countries.
\end{hypothesis}

While \hyp{cfr} ignores medical, demographic and other differences
between countries, I believe it unlikely that the CFR is very
different across different countries. In the end, its the same type of
virus spreading in all countries. This suggests that differences in
estimated CFRs simply reflect differences in the ability of countries
to actually observe all infected individuals, i.e. due to more or less
effective tracking and testing procedures. To illustrate this effect,
a true CFR of $1\%$ is assumed in the following. This is consistent
with current knowledge and had also been used in other studies
\cite{imperial2}. Just from the estimated values any CFR below the
minimum of all estimates (about $2\%$ found for Austria and South
Korea) and above $0.1\%$ (which would imply an observed fraction above
one for Belgium) is compatible with the data.

\begin{figure}
  \includegraphics[width=1\textwidth]{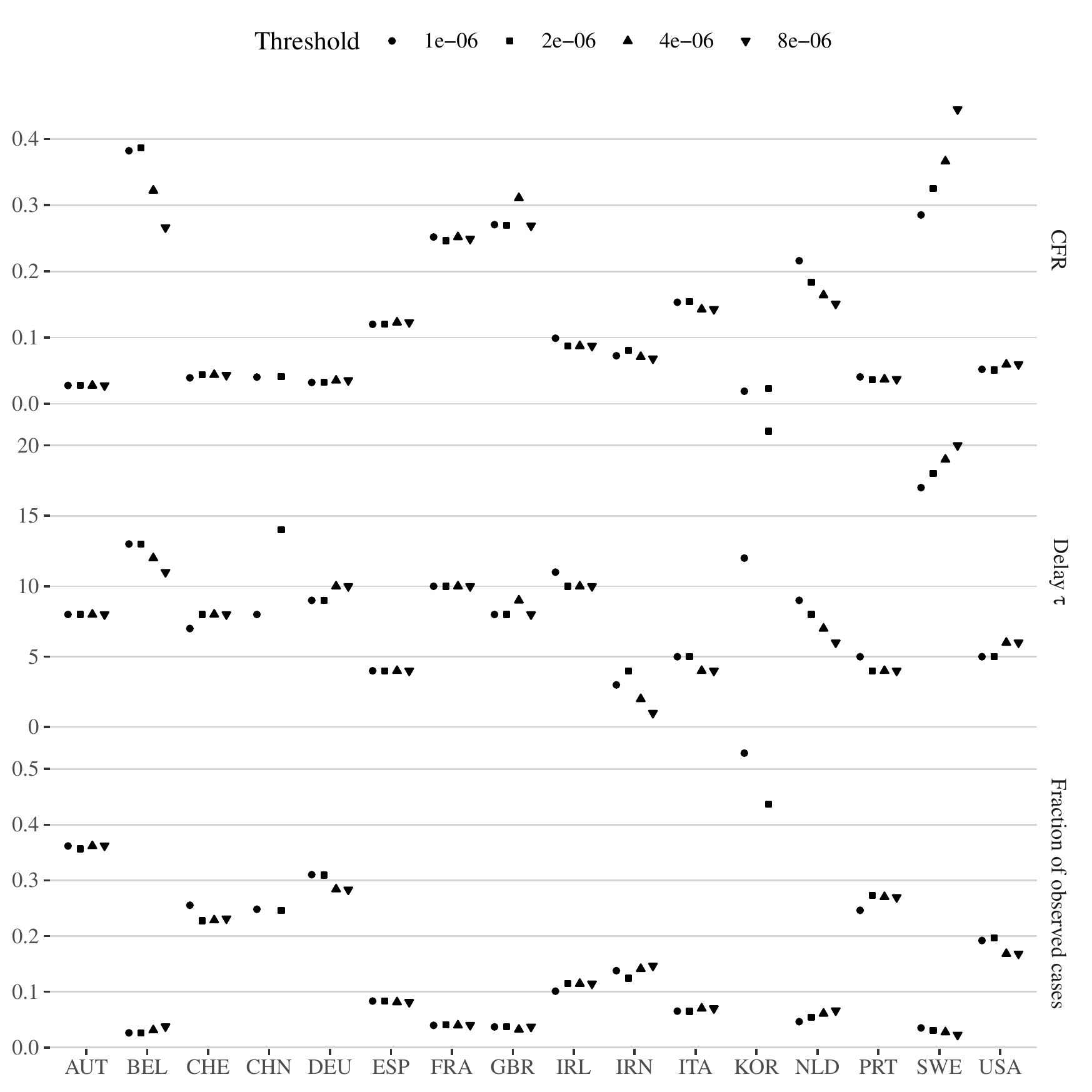}
  \caption{\label{fig:estimates}}
\end{figure}

\begin{figure}
  \includegraphics[width=1\textwidth]{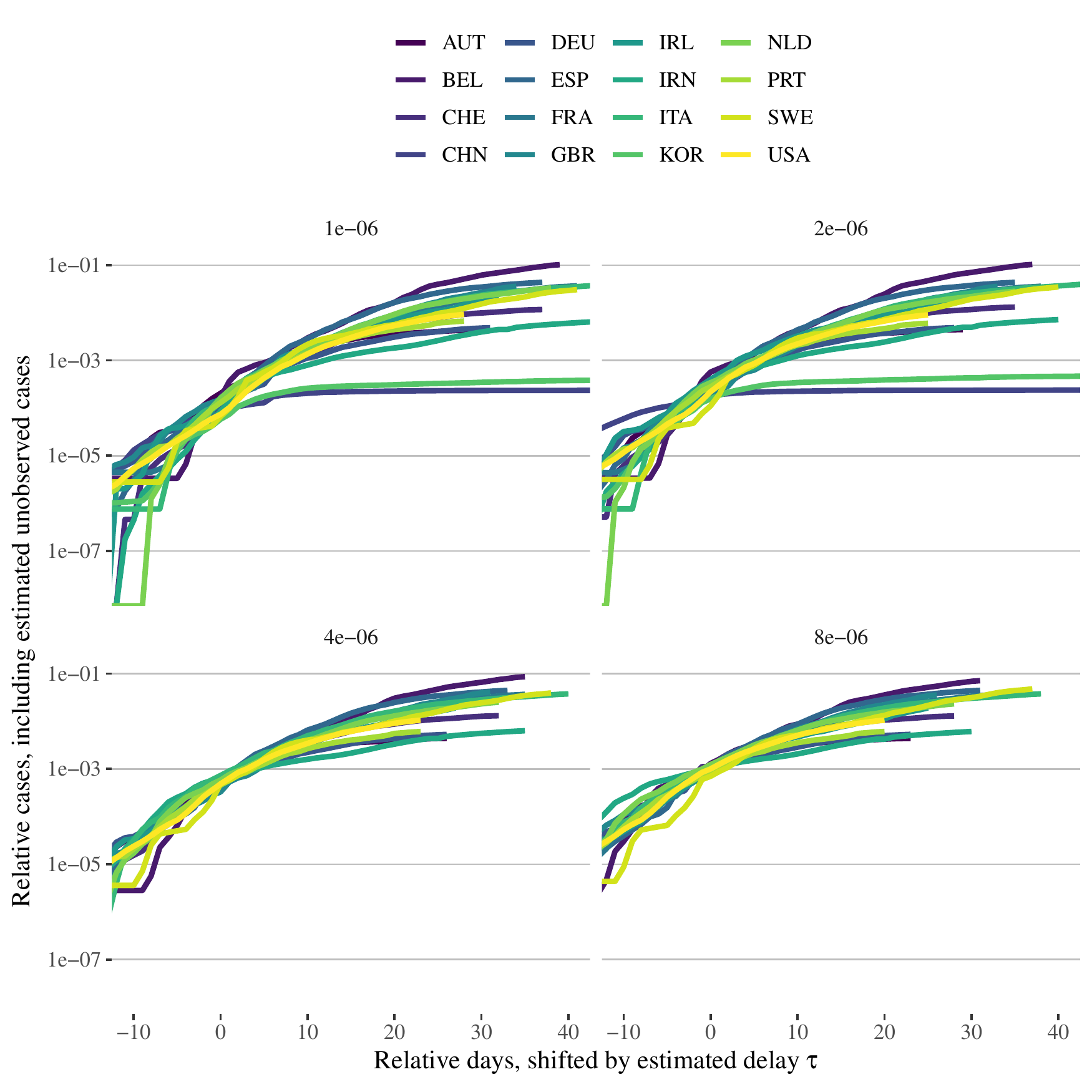}
  \caption{\label{fig:aligned_case_est}}
\end{figure}

\fig{estimates} shows the country specific estimates of reporting
delay, CFR and fraction of observed cases (assuming a true CFR of
$1\%$) obtained in this fashion. In turn, \fig{aligned_case_est} shows
the implied relative case counts when shifted by the estimated delays
and scaled to reflect the unobserved fraction of cases for each
country. Notably, these implied counts all align nearly as good as the
death counts in \fig{aligned_data} (right panel) even though the
initial threshold was based on the deaths counts alone. The
supplementary \fig{scaling_case_est} shows that this holds also when
re-scaling time according to the growth rate of deaths. Overall, the
collapse of implied case dynamics convincingly illustrates that the
relation between case and death counts is fully and reliably captured
by two parameters -- compatible with three reasonable assumptions.

\section{Discussion}

In reality, an additional delay between an infection and its
corresponding positive test result can be assumed. Therefore, the
fraction of observed cases will be even lower than obtained by the
analysis above. Unfortunately, assuming a sufficiently flexible model
for the growth of the actual cases already the CFR and the fraction of
observed cases, let alone an additional delay, are not jointly
identifiable.

\subsection{Epidemic modeling}
The basic SIR model \cite{Newman}, assumes that an infection unfolds when
susceptible (S) individuals become infected (I) -- which in turn
infect further susceptible individuals. Finally, infected individuals
recover (R) (or die) and are no longer susceptible. In continuous
time, the dynamics can be described by the following system of
ordinary differential equations (ODEs):
\begin{align*}
  \frac{dS}{dt} &= - \beta \frac{I_t}{N} S_t \\
  \frac{dI}{dt} &= \beta \frac{I_t}{N} S_t - \gamma I_t \\
  \frac{dR}{dt} &= \gamma I_t
\end{align*}
where $N \equiv S_t + I_t + R_t$ is constant over time. Model
parameters are
\begin{itemize}
\item the infection rate $\beta$
\item and the recovery rate $\gamma$.
\end{itemize}
In this model, the average time of infection is $\gamma^{-1}$ giving
rise to a {\em basic reproduction number} of $R_0 = \beta
\gamma^{-1}$.

SIR models and extensions are widely used in epidemic modeling. The
have also been applied to the understand the dynamics of the ongoing
Covid-19 pandemic
\cite{arxiv:2002.07572,arxiv:2004.01105,10.1126/science.abb3221,https://www.medrxiv.org/content/10.1101/2020.02.27.20028639v2}.
In particular, models including the possibility of unobserved cases or
including a reporting delay have been developed. Within the SIR
framework, both effects can be included in several ways, most easily
by assuming that observed cumulative infections are simply a fraction
$\alpha \in [0, 1]$ of previous total infections $I_t + R_t$,
i.e. $\alpha (I_{t - \tau} + R_{t - \tau})$. A more elaborate attempt
instead considers more detailed dynamics of the form
\begin{align*}
  \frac{dS}{dt} &= - \beta_I \frac{S_t}{N} I_t - \beta_O \frac{S_t}{N} O_t - \beta_U \frac{S_t}{N} U_t \\
  \frac{dI}{dt} &= \beta_I \frac{S_t}{N} I_t + \beta_O \frac{S_t}{N} O_t + \beta_U \frac{S_t}{N} U_t - \gamma_I I_t \\
  \frac{dO}{dt} &= \alpha \gamma_I I_t - \gamma_R O_t \\
  \frac{dU}{dt} &= (1 - \alpha) \gamma_I I_t - \gamma_R U_t \\
  \frac{dR}{dt} &= \gamma_R (O_t + U_t)
\end{align*}
where a fraction $\alpha$ of infected individuals $I_t$ is observed
($O_t$) after an initial delay $\frac{1}{\gamma_I}$. In any case,
whether observed or not, individuals recover (or die) after an
additional delay. In general, the infection rates $\beta_I, \beta_O,
\beta_U$ could be different for initial infections and observed vs
unobserved cases\footnote{An effective quarantine would be modeled via
  $\beta_O \equiv 0$.}.

\begin{figure}
  \includegraphics[width=0.45\textwidth]{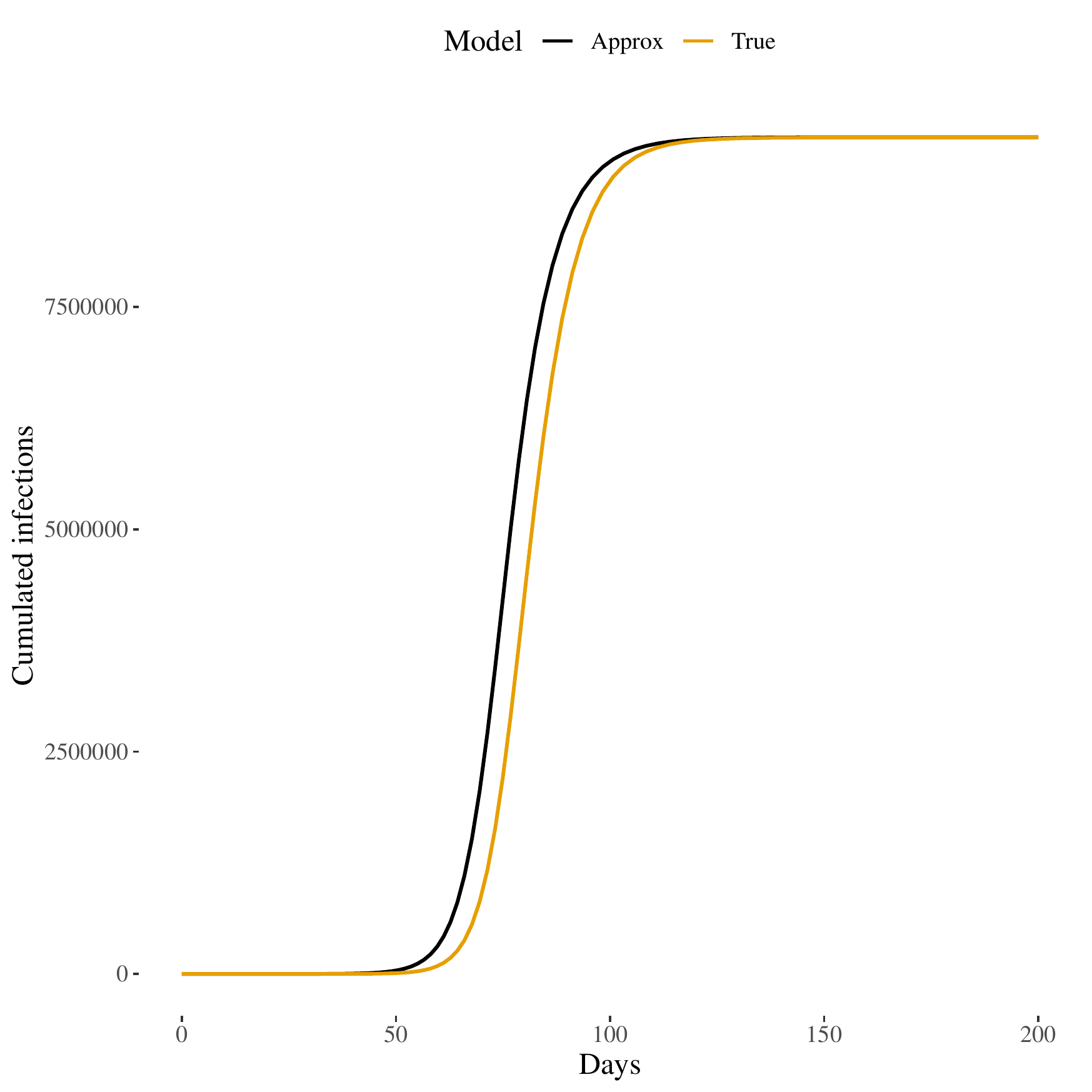}
  \includegraphics[width=0.45\textwidth]{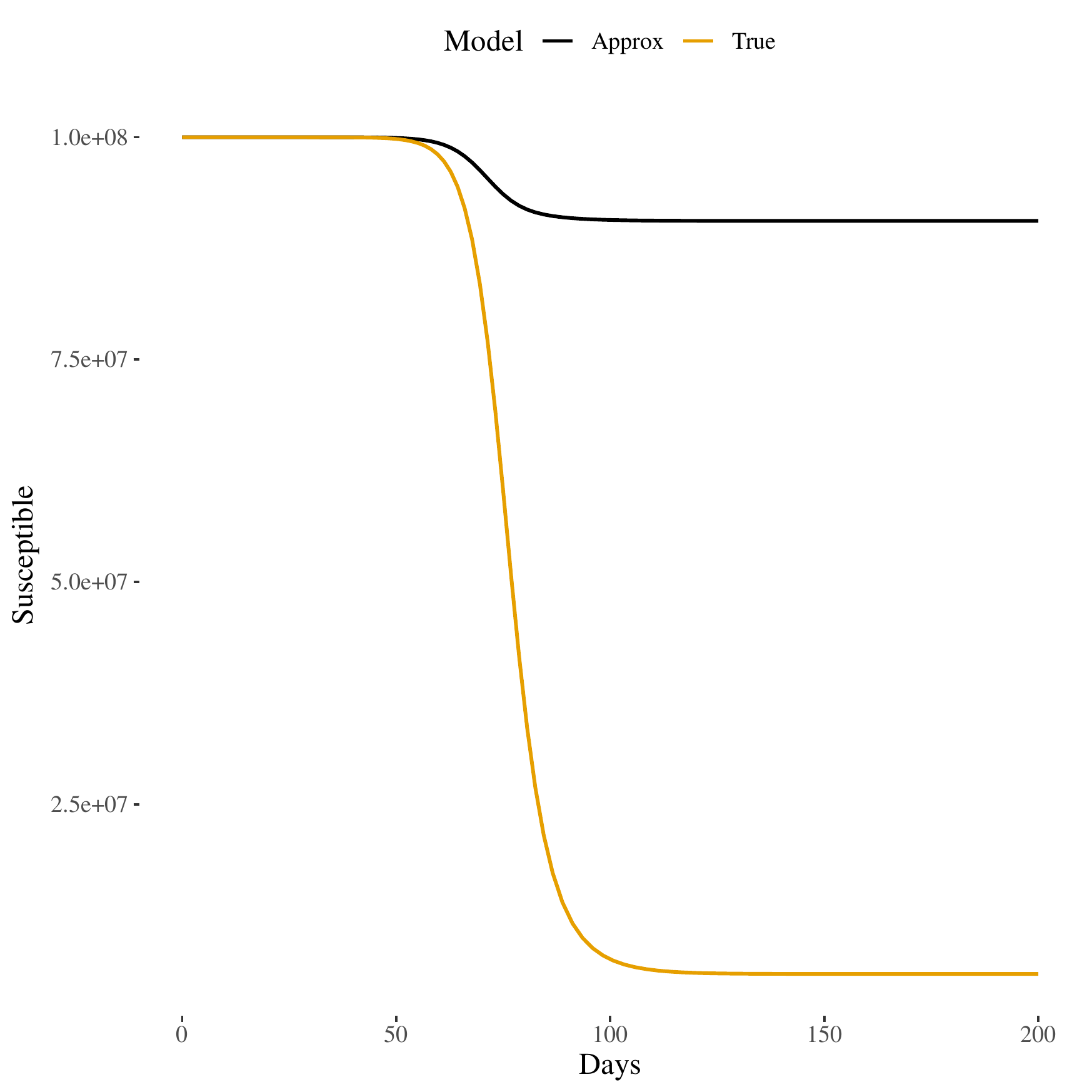}
  \caption{\label{fig:SIRapprox} Total cumulative observed infections
    and number of susceptible individuals in two simulated model with
    observation fractions $\alpha = 0.1$ (true), $\alpha' = 1$ and
    time shift $\tau = 5$ (approx). In the second model, the epidemic
    is stopped due to mitigation measures which are modeled via
    $\beta'(t)$ as explained in the main text.}
\end{figure}

In addition, mitigation measures, e.g. social distancing, can be
easily included by assuming that $\beta$'s are functions of
time. E.g. \cite{arxiv:2004.01105} assumes one or several (soft) step
functions where $\beta$ drops after measures have been
implemented. Unfortunately, as we show now a model including a
time-varying $\beta$ as well as unobserved cases is not identifiable.
For simplicity, consider the above model with $\beta_I = \beta_O =
\beta_U =: \beta$. Then, new infections arise with intensity $\beta
\frac{S_t}{N} (I_t + O_t + U_t)$ which in turn translate into observed
cases with intensity $\alpha \gamma_I I_t$. Now assume a second model
with $\alpha' = 1 > \alpha$ which nevertheless exhibits the same
dynamics with an additional time shift $\tau$. By using a time varying
$\beta'(t)$ such that
\begin{align*}
  \beta'(t) &= \alpha \beta \frac{S_{t + \tau}}{S'_t}
\end{align*}
we obtain exactly the same number of observed cases, i.e. $O'_{t -
  \tau} = O_{t}$. Note that as $\alpha' > \alpha$, we have that
$S_{t} < S'_{t-\tau}$ and $S_t$ is a sigmoidal function of time due to
the SIR dynamics. Furthermore, when the population is large, i.e. $N
\gg 1$ and $S_0 \approx N$ the resulting $\beta'(t)$ is mostly driven
by the drop in $S_{t+\tau}$ as compared to the much smaller change in
$S'_t$. Indeed, \fig{SIRapprox} shows the dynamics of the above model
with $\beta = 0.3, \gamma_I = \gamma_R = \frac{2}{10}$\footnote{Giving
  rise to an $R_0$ of $3$.}, $\alpha = 0.1$ starting from $(N = 10^8,
1, 0, 0, 0)$. In turn, assuming $\alpha' = 1$ and $\tau = 5$, the time
varying infectivity $\beta'(t)$ is approximated by the best-fitting
logistic sigmoid of the form $\beta_1 + (\beta_2 - \beta_1)
\sigma(\frac{t - \tau}{T})$. Note that the number of observed cases is
identical, just shifted by $\tau$, whereas the final fraction of susceptible
individuals is vastly different. Indeed, in the first case the
epidemic is stopped by group immunity whereas in the second case
effective mitigation measures are imposed. Correspondingly, police
implications would be vastly different in the two situations even
though they are observationally indistinguishable.

\subsection{Implications}

Instead of detailed modeling of epidemic dynamics, which is further
complicated due policy actions requiring flexible models with
delicately chosen parameters, the present analysis is based on visual
inspection of the reported data. Overall, relative case and deaths
counts (observed for country $c$) seem to be related as follows:
\begin{align*}
  d^c_t &= \mathrm{cfr} \cdot a^c_{t - {\tau^c}} \\
  c^c_t &= \alpha^c \cdot a_t
\end{align*}
where $a^c_r$ denotes the actual infections a fraction $\alpha^c \in
[0, 1]$ is observed. A suitable reporting delay $\tau^c$ can be
estimated by visual inspection of the data, but again the fraction of
observed cases $\alpha^c$ and CFR $\mathrm{cfr}$ are not jointly
identifiable if there exist sets of parameters such that $a'_{t -
  {\tau}} = \alpha a_t$, as is the case for dynamic SIR type
models. In the end, any epidemic modeling implicitly or explicitly
chooses a parametric form for the latent growth process $a_t$ and will
not be identified if sufficiently flexible. Yet, assumption three of
a constant CFR across all countries allows to derive
\begin{enumerate}
\item a range of values consistent among all countries,
\item as well as recover the corresponding fraction of observed cases
  in each country.
\end{enumerate}

\begin{figure}
  \includegraphics[width=1\textwidth]{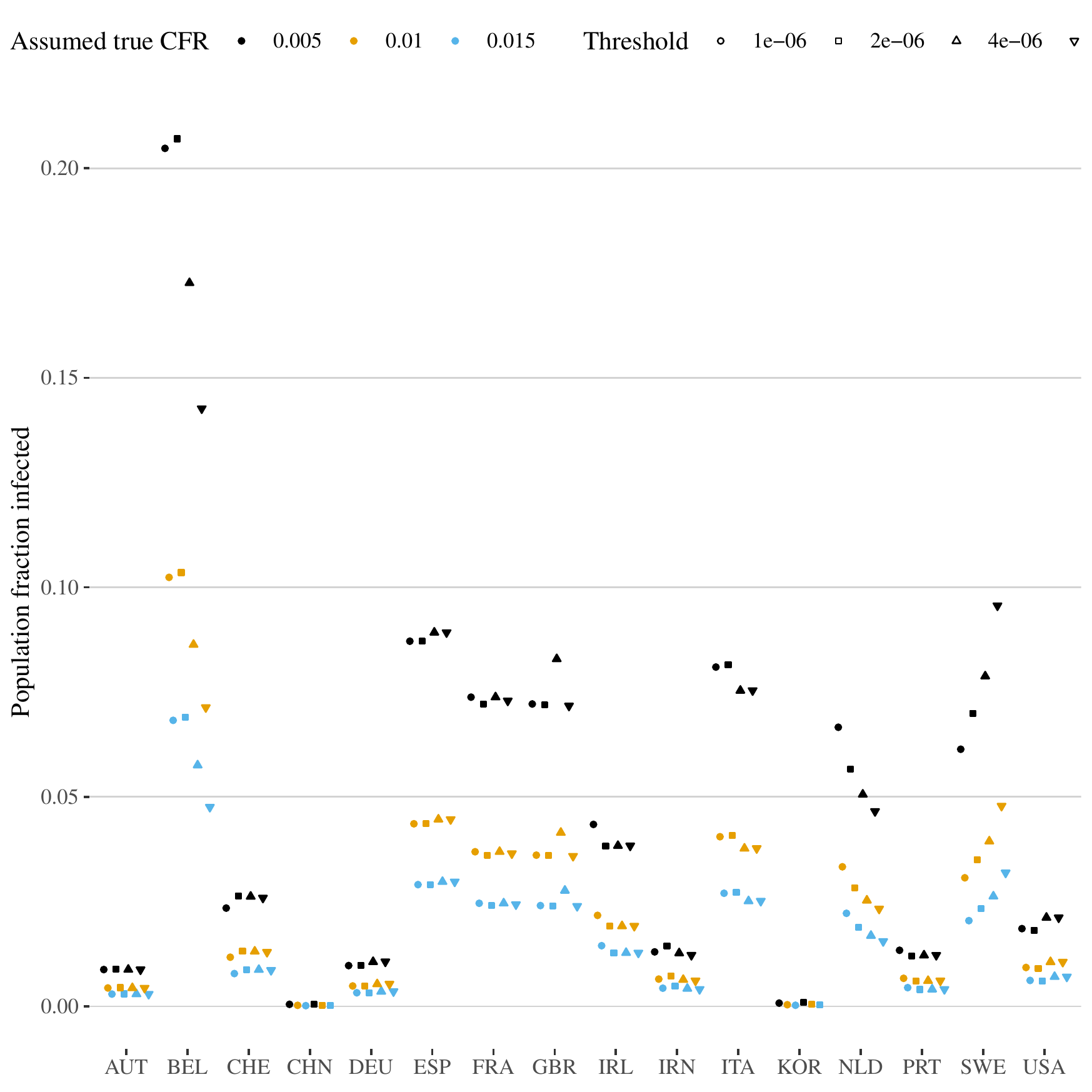}
  \caption{\label{fig:est_pop_frac} Estimated fraction of population
    already infected assuming a reasonable range of true CFRs.}
\end{figure}

Thereby, assuming a reasonable true CFR value, i.e. from the model
implied range $0.1\%$ to $2\%$ which is also consistent with current
knowledge, and using the estimated delay, the actual case numbers can
be reconstructed. \fig{est_pop_frac} shows the resulting actual
relative infection counts across several countries. Note that despite
the simplicity of this analysis, the estimated numbers compare
favorable \cite{imperial2}. Indeed, I would rather trust these even
more as they do not rely on complex modeling assumptions but follow
from visual inspection of the data.

Overall, I have illustrated that much of the variability between
observed case and deaths counts between different countries can be
explained by two parameters. Namely, the reporting delay $\tau$ and
the fraction of observed cases. Especially the reporting delay
exhibits crucial differences between countries and needs to be taken
into account when comparing data and planning actions. In particular,
containment is challenging when long incubation times are involved
\cite{https://doi.org/10.1101/2020.03.25.20043109} but a combination
of case tracing and isolation policies could be effective
\cite{fraser04,kubinec}. Thus, detailed epidemic modeling is certainly
needed in order to judge the effectiveness of current mitigation
measures across different countries \cite{imperial2,arxiv:2004.01105}.
On the other hand, important parameters need to fixed based on
additional knowledge as they cannot be identified within sufficiently
flexible models. In the end, data analysis and detailed modeling alone
only gets us only that far and more extensive testing is urgently
needed to obtain reliable knowledge about the current progression of
the Covid-19 pandemic.

\bibliographystyle{abbrv}
\bibliography{notes}

\clearpage

\appendix
\renewcommand\appendixname{Supplement}
\beginsupplement

\section{Data collapse by re-scaling time}

\begin{figure}
  \includegraphics[width=1\textwidth]{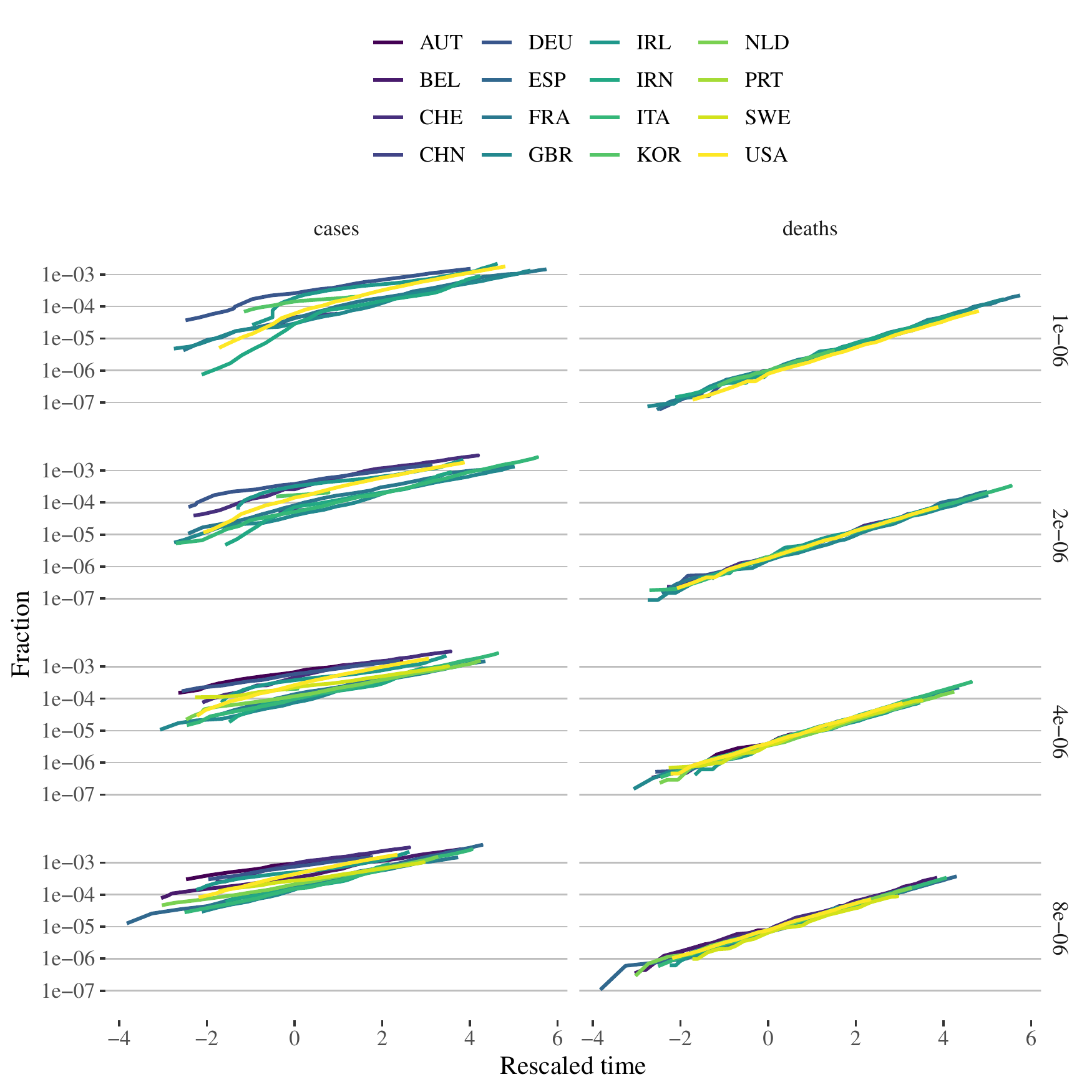}
  \caption{\label{fig:scaling} Aligned data as in \fig{aligned_data},
    but time is additionally re-scaled to match local growth rate of
    the epidemics.}
\end{figure}

Aligning the data as in \fig{aligned_data} still shows
country-specific differences in the temporal course of epidemic
spreading. Much of this difference can be attributed to the speed at
which the epidemic spreads in different countries. Estimating the
local growth rate of deaths $\frac{d\log d_t}{dt}$ by the three day
running average of observed changes $\log d_{t+1} - \log d_t$,
relative time, i.e. relative to the threshold of total deaths reached,
is re-scaled to match local growth rates. \fig{scaling} shows the
resulting data collapse for $d_t$ and the corresponding $c_t$ dynamics.

Further, taking the estimated relation between cases and deaths via
CFR and country specific delays into account an almost complete data
collapse for the cases is obtained. Not that as in the main text, data
are aligned according to relative death counts only. Furthermore, the
temporal re-scaling is based on the estimated growth rate from the
death counts as well. Yet, shifting and scaling case data according to
the estimated country specific delay and fraction of observed cases
leads to an almost complete data collapse as well.
\begin{figure}
  \includegraphics[width=1\textwidth]{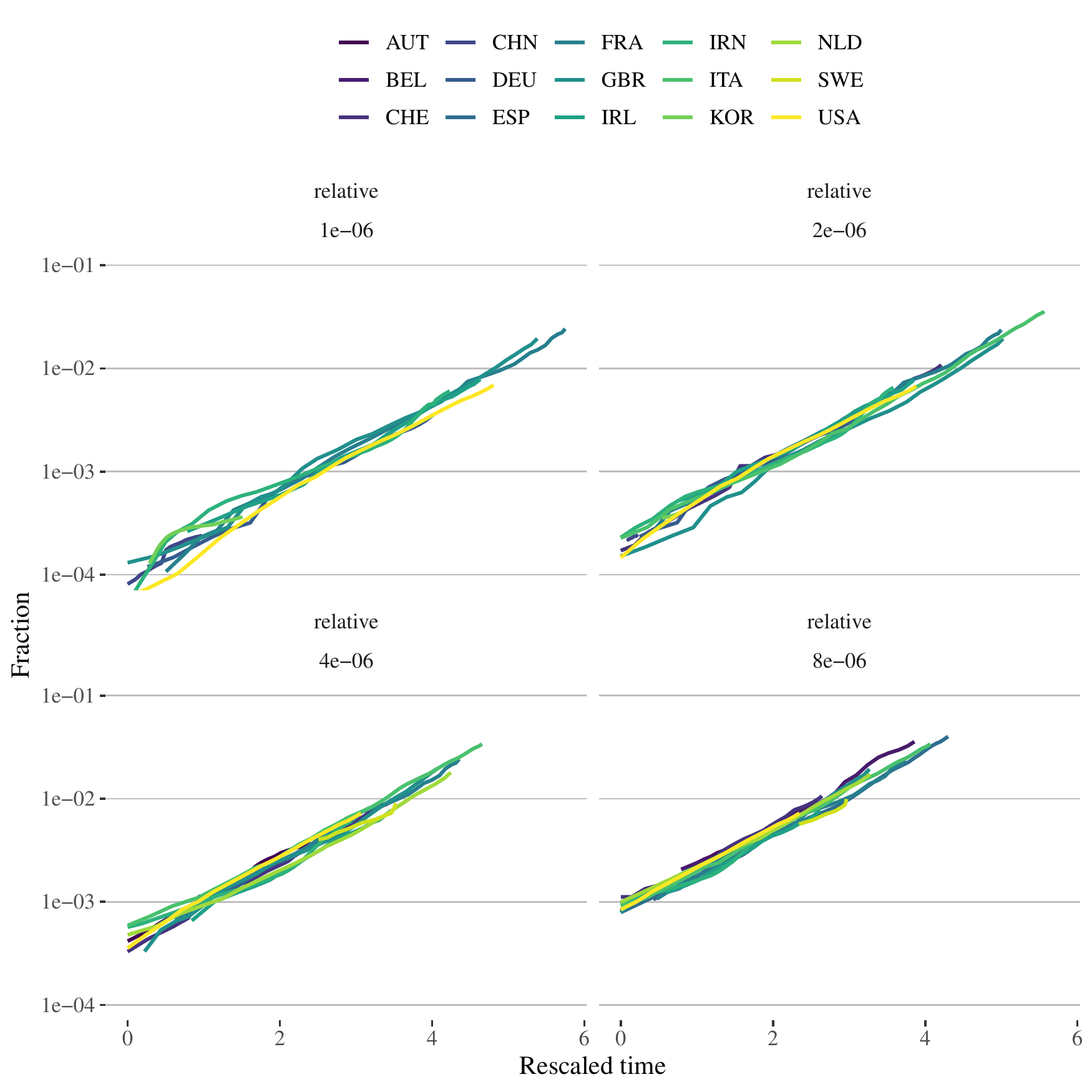}
  \caption{\label{fig:scaling_case_est} As \fig{aligned_case_est}, but
    time is additionally re-scaled to match local growth rate of the
    epidemics.}
\end{figure}

\section{NY Times style figures}

As individual countries can be hard to identify in Figures
\ref{fig:aligned_data} and \ref{fig:aligned_case_est}, the NY Times
featured panel views where each country is highlighted above a
background of all countries. Here, I provide similar figures for
relative death and case counts using a threshold of two deaths per
million inhabitants.
\begin{figure}
  \includegraphics[width=1\textwidth]{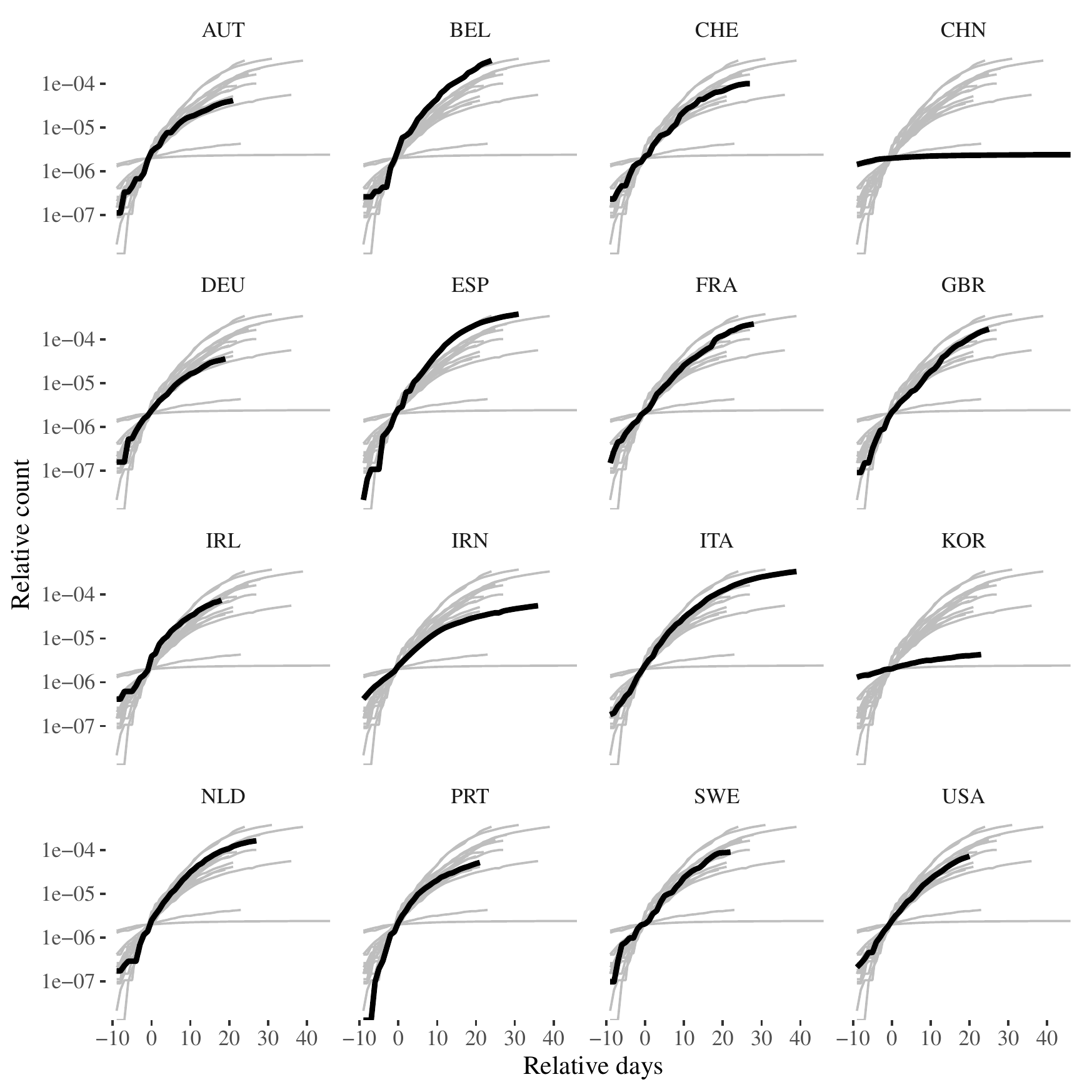}
  \caption{\label{fig:align_nyt} Details of aligned relative death
    counts for threshold of two deaths per million.}
\end{figure}

\begin{figure}
  \includegraphics[width=1\textwidth]{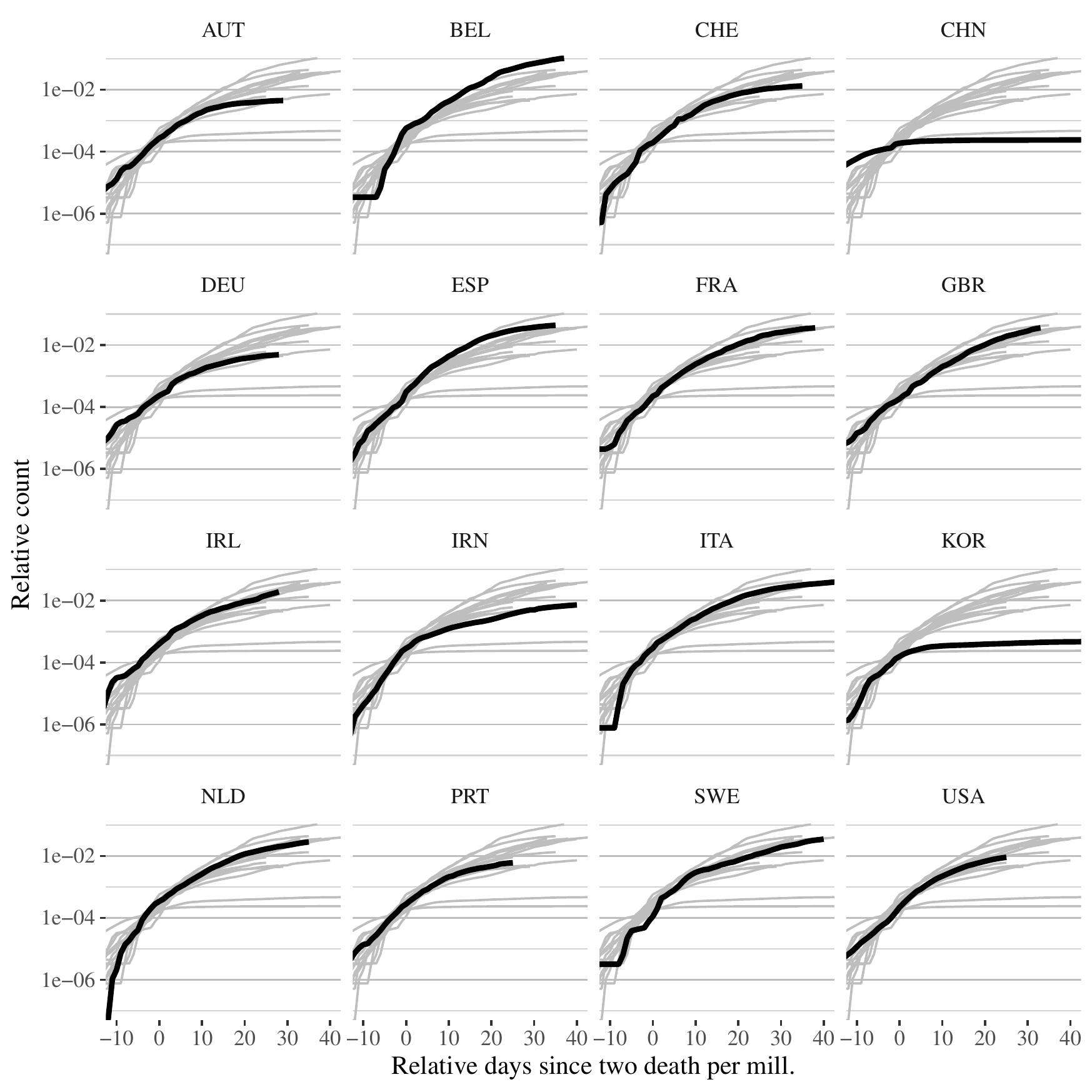}
  \caption{\label{fig:align_estcase_nyt} Details of aligned and
    adjusted case counts for threshold of two deaths per million.}
\end{figure}

\section{Uncertainty estimates from SIR model}

\begin{figure}
  \begin{minipage}{0.66\textwidth}
    \includegraphics[width=1\linewidth]{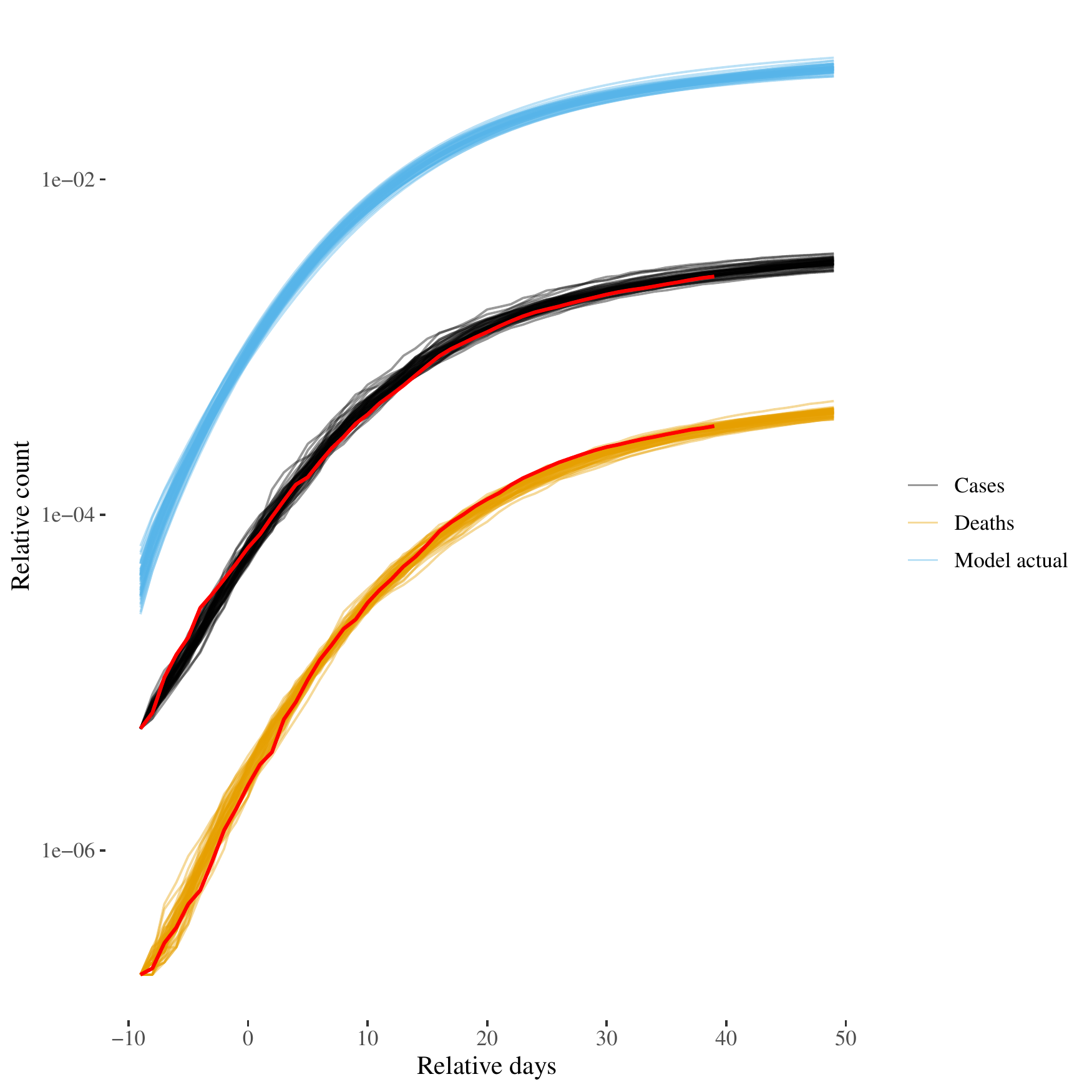}      
  \end{minipage}
  \begin{minipage}{0.33\textwidth}
    \includegraphics[width=1\linewidth]{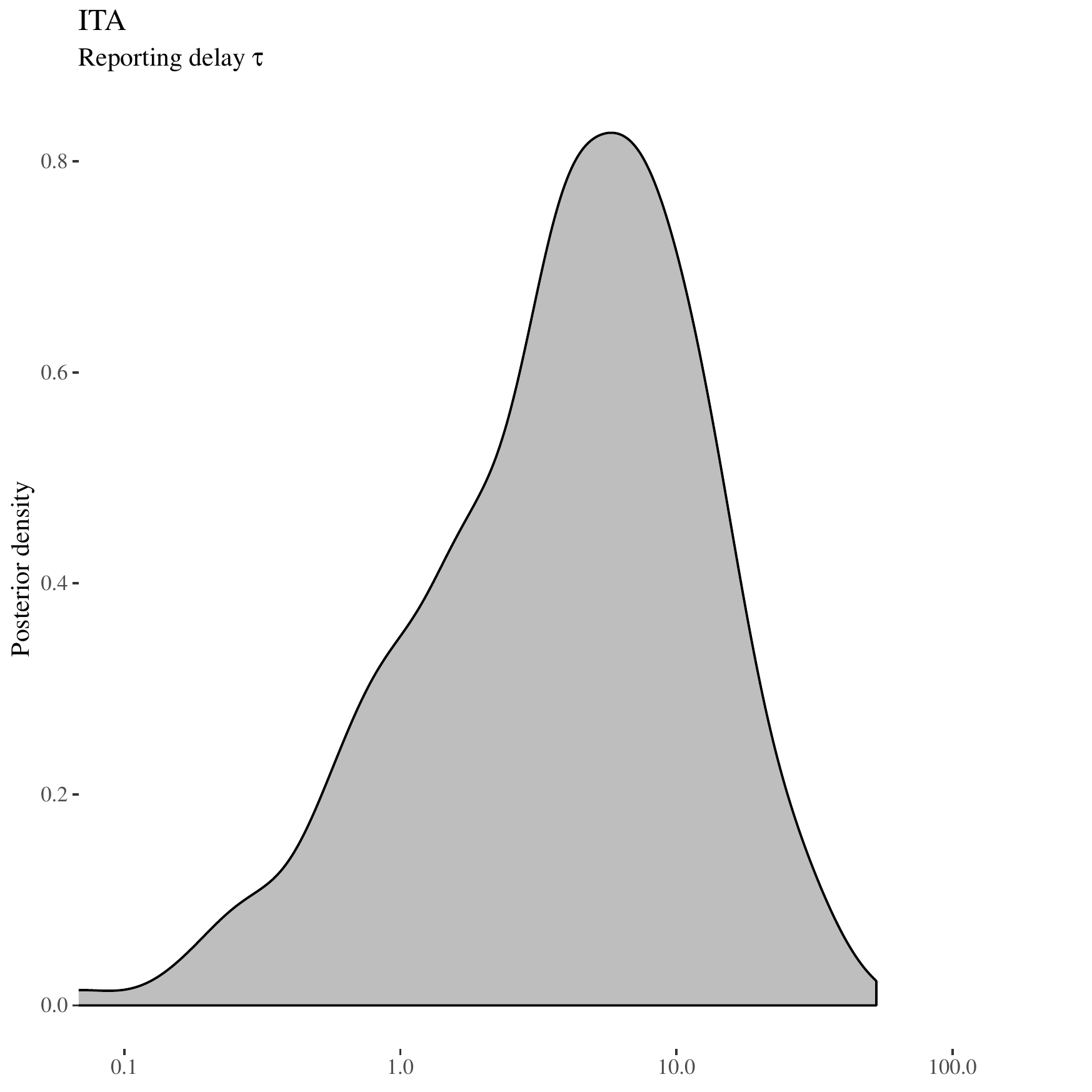}
    \includegraphics[width=1\linewidth]{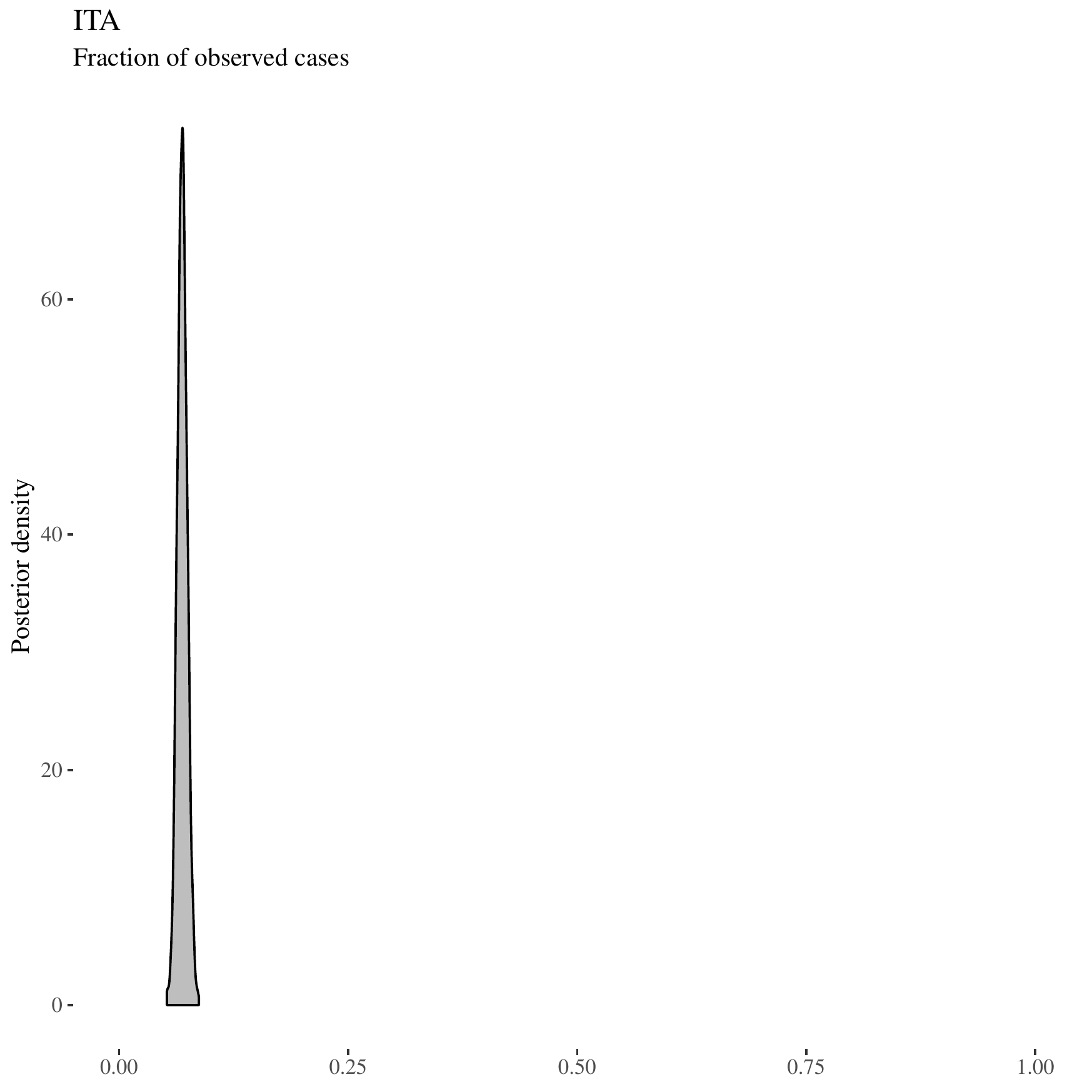}
  \end{minipage}
  \begin{minipage}{0.66\textwidth}
    \includegraphics[width=1\linewidth]{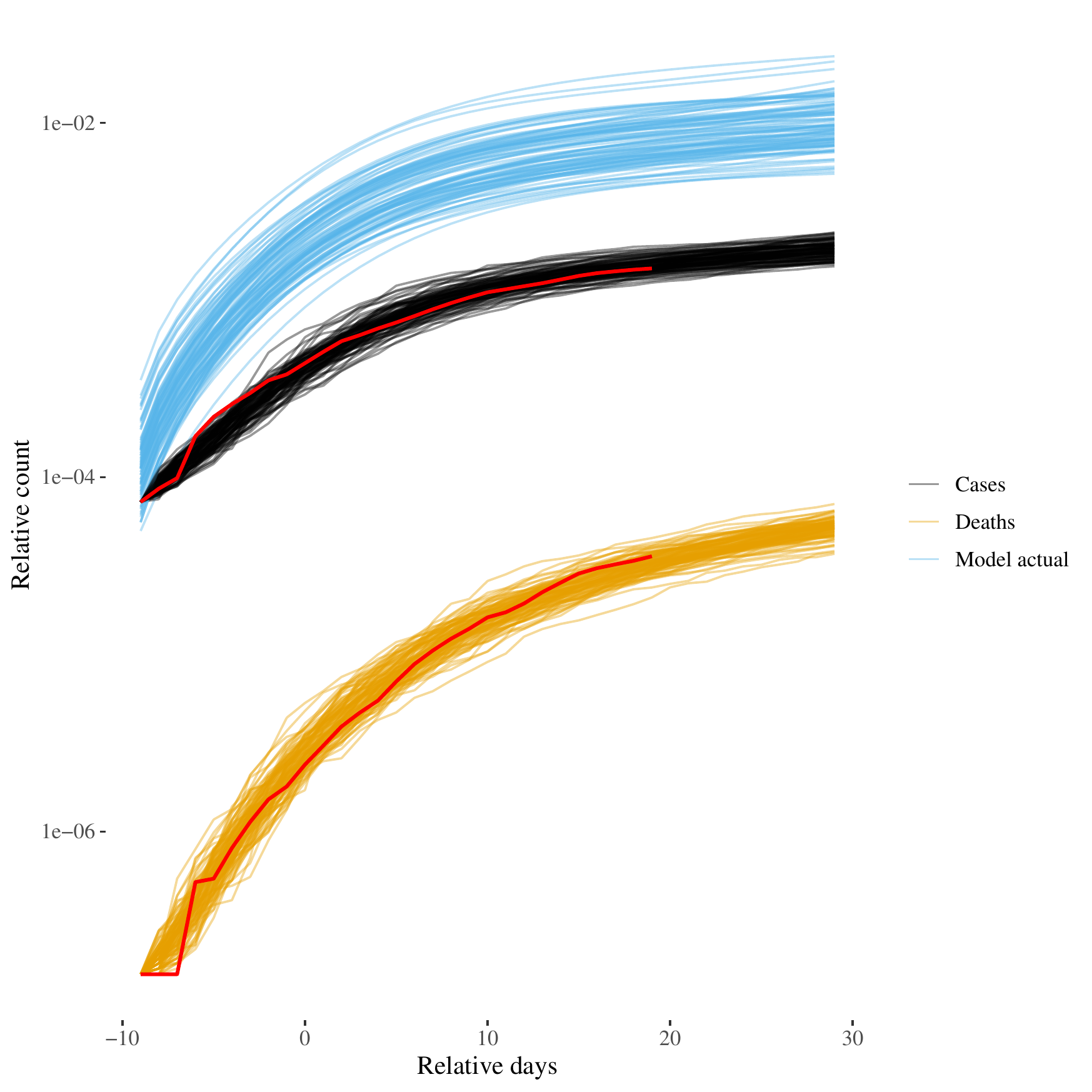}      
  \end{minipage}
  \begin{minipage}{0.33\textwidth}
    \includegraphics[width=1\linewidth]{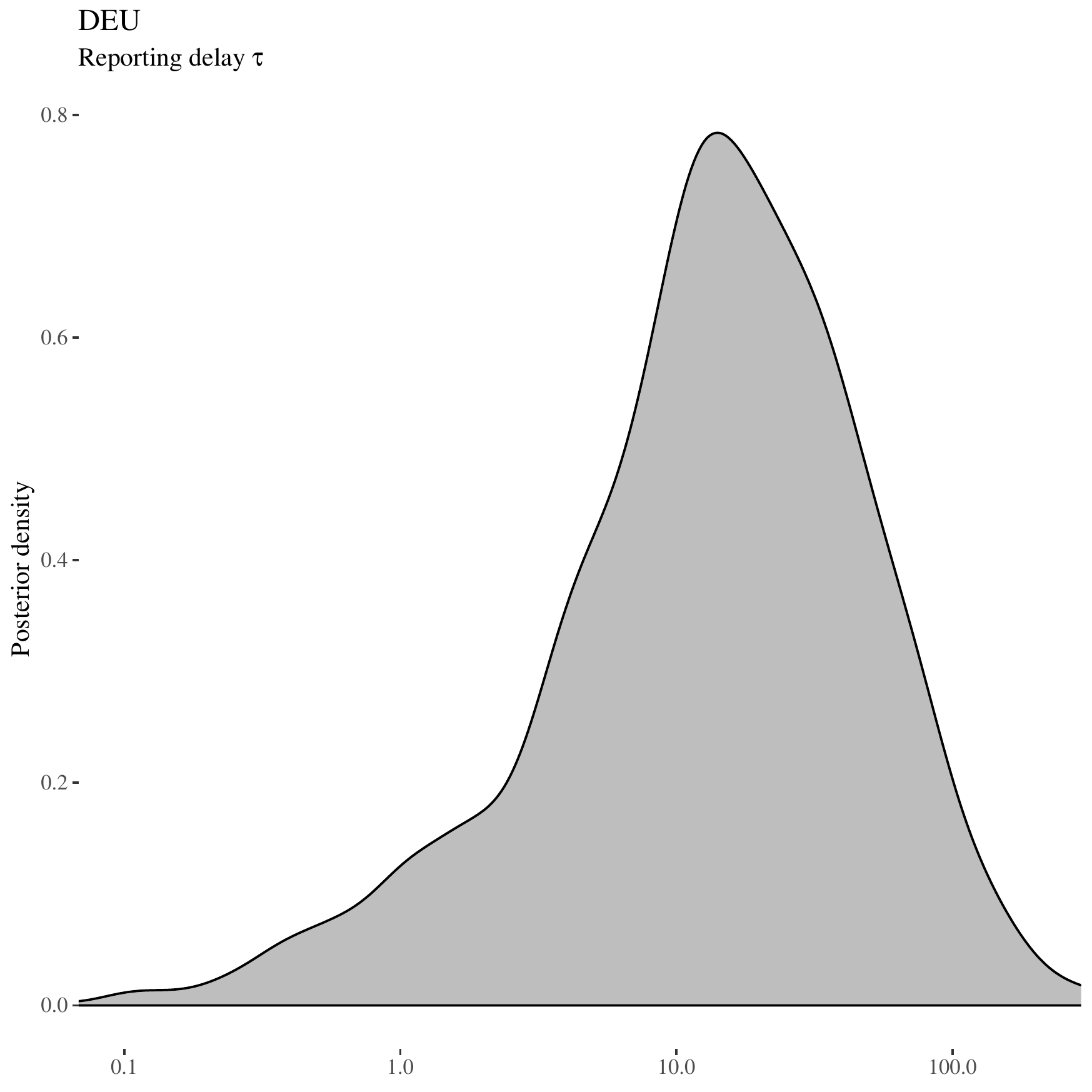}
    \includegraphics[width=1\linewidth]{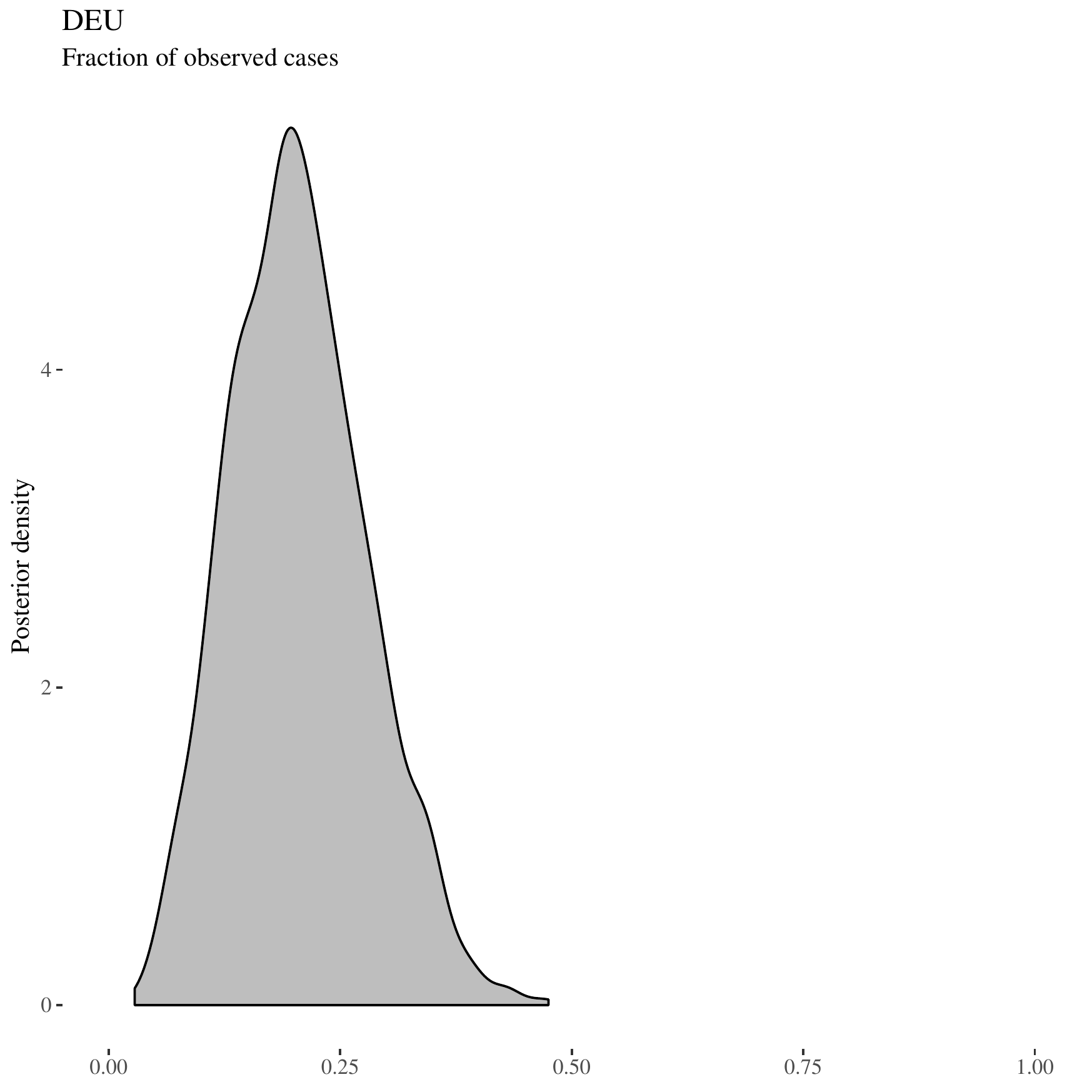}
  \end{minipage}
  \caption{\label{fig:sir_fit} Model predictions and estimated
    parameters from SIR model fitted to data from Italy (top) and
    Germany (bottom).}
\end{figure}

Note that an SIR model already includes a natural delay between
infections and recovery (or death). Indeed, the total number of cases
is given by $C_t = I_t + R_t$ while the cumulative death toll is
obtained as $\mathrm{cfr} R_t$, i.e. modeling that a fraction of
individuals does not recover but dies instead. Assuming that only a
fraction $\alpha$ of cases is observed, the model is estimated with
the following sampling distribution
\begin{align*}
  C^{\mathrm{obs}}_{t+1} - C^{\mathrm{obs}}_t &\sim \mathrm{NegativeBinomial}\left(\alpha \frac{dC^{\mathrm{model}}_t}{dt}, \phi_C \right) \\
  D^{\mathrm{obs}}_{t+1} - D^{\mathrm{obs}}_t &\sim \mathrm{NegativeBinomial}\left(\mathrm{cfr} \frac{dR^{\mathrm{model}}_t}{dt}, \phi_D \right) \; .
\end{align*}
Thus, observed daily changes are related to the model implement
changes via an over-dispersed Poisson aka negative binomial
distribution. \fig{sir_fit} shows the resulting estimates assuming
$\beta_t = \beta_1 + (\beta_2 - \beta_1) \sigma(\frac{t - \tau}{T})$
and $\mathrm{cfr} = 1\%$\footnote{Due to the non-identifiability
  derived in the main text either $\alpha$ or $\mathrm{cfr}$ needs to
  be fixed.}.  The SIR model assuming a single change point in the
infectivity, via the logistic sigmoid $sigma(\cdot)$ in $\beta_t$
reflecting the implementation of social distancing is clearly able to
capture the epidemic dynamics. Yet, parameter uncertainties,
especially about the reporting delay can be large\footnote{The high
  uncertainty could also reflect that an SIR dynamics is misspecified
  in that it corresponds to an exponential delay distribution. Such
  additional model assumptions need to be carefully chosen in order to
  obtain meaningful parameter estimates.}.

Bayesian estimates have been carried out using {\em Stan} (full code
available from my
\href{Github}{\url{https://github.com/bertschi/Covid}} repository) and
using weakly informative broad normal or student-t prior distributions
on all parameters.

\end{document}